\documentclass{aastex631}
\usepackage{amsmath}

\newcommand{\g}{\gamma}

\begin{document}

%\title{An Updated Measurement of the Extragalactic Background Light Using GeV $\gamma$ rays}

\title{A New Measurement of the Extragalactic Background Light using 15\,yr of {\it Fermi}-Large Area Telescope Data}

\author[0000-0001-7796-8907]{Anuvab Banerjee}
\affiliation{Department of Physics and Astronomy, Clemson University, Clemson, SC, 29631}

\author[0000-0001-5941-7933]{Justin D.\ Finke}
\affil{U.S.\ Naval Research Laboratory, Code 7653, 4555 Overlook Ave.\ SW, Washington, DC, 20375-5352, USA }

\author[0000-0002-6584-1703
]{Marco Ajello}
\affiliation{Department of Physics and Astronomy, Clemson University, Clemson, SC, 29631}

%\collaboration{20}{(AAS Journals Data Editors)}

\author[0000-0002-3433-4610]{Alberto Dom\'{i}nguez}
\affiliation{IPARCOS and Department of EMFTEL, Universidad Complutense de Madrid, E-28040 Madrid, Spain}

\author[0000-0001-7405-9994]{Abhishek Desai}
\affiliation{NASA Postdoctoral Program Fellow, NASA Goddard Space Flight Center, Greenbelt, MD, 20771, USA}

\author[0009-0004-9545-794X]{Joshua Baxter}
\affiliation{Institute for Cosmic Ray Research, University of Tokyo, Kashiwa, Chiba 277-8582, Japan}

\author[0000-0002-8028-0991]{Dieter Hartmann}
\affiliation{Department of Physics and Astronomy, Clemson University, Clemson, SC, 29631}

\author[0000-0001-7774-5308]{Vaidehi  S. Paliya}
\affiliation{Inter-University Centre for Astronomy and Astrophysics (IUCAA), SPPU Campus, Pune, 411007, Maharashtra, India}

%\affiliation{AAS Journals Associate Editor-in-Chief}

%\author{Amy Hendrickson}
%\altaffiliation{AASTeX v6+ programmer}
%\affiliation{TeXnology Inc.}

%% Note that the \and command from previous versions of AASTeX is now
%% depreciated in this version as it is no longer necessary. AASTeX 
%% automatically takes care of all commas and "and"s between authors names.

%% AASTeX 6.31 has the new \collaboration and \nocollaboration commands to
%% provide the collaboration status of a group of authors. These commands 
%% can be used either before or after the list of corresponding authors. The
%% argument for \collaboration is the collaboration identifier. Authors are
%% encouraged to surround collaboration identifiers with ()s. The 
%% \nocollaboration command takes no argument and exists to indicate that
%% the nearby authors are not part of surrounding collaborations.

%% Mark off the abstract in the ``abstract'' environment. 
\begin{abstract}

The extragalactic background Light (EBL) from ultraviolet to infrared comprises the emission from all stars, galaxies, and actively accreting black holes in the observable Universe. A precise measurement of the EBL is critically important to probe models of star formation and galaxy evolution. The EBL  can be measured via the absorption imprint left on the spectra of gamma-ray blazars. In this work, we rely on 15 years of {\it Fermi}-LAT data and 1576 blazars to measure the EBL optical depth in the $0<z<4.3$ range. We detect the EBL attenuation with $\sim23\sigma$ significance and measure the optical depth in 19 redshift bins, extending the coverage and improving on our previous results. This allows us to reconstruct the EBL evolution and find general consistency with recent EBL models. These results represent the most precise determination of the EBL with GeV $\gamma$ rays to date.

\end{abstract}

%% Keywords should appear after the \end{abstract} command. 
%% The AAS Journals now uses Unified Astronomy Thesaurus concepts:
%% https://astrothesaurus.org
%% You will be asked to selected these concepts during the submission process
%% but this old "keyword" functionality is maintained in case authors want
%% to include these concepts in their preprints.

%% From the front matter, we move on to the body of the paper.
%% Sections are demarcated by \section and \subsection, respectively.
%% Observe the use of the LaTeX \label
%% command after the \subsection to give a symbolic KEY to the
%% subsection for cross-referencing in a \ref command.
%% You can use LaTeX's \ref and \label commands to keep track of
%% cross-references to sections, equations, tables, and figures.
%% That way, if you change the order of any elements, LaTeX will
%% automatically renumber them.
%%
%% We recommend that authors also use the natbib \citep
%% and \citet commands to identify citations.  The citations are
%% tied to the reference list via symbolic KEYs. The KEY corresponds
%% to the KEY in the \bibitem in the reference list below. 

\section{Introduction}
\label{intro}

The Extragalactic Background Light (EBL) is the integrated background emission at ultraviolet (UV), optical and infrared (IR) wavelengths contributed by stars and dust in galaxies within the observable universe {\citep[e.g.,][]{dwek2013APh}}. The local ($z=0$) EBL spectral energy distribution (SED) shows a double-peaked profile \citep{driver2016measurements}, with the {shorter wavelength} peak produced by direct stellar emission, while the {longer wavelength} peak is produced by {galactic starlight that is reprocessed by dust harbored within the same galaxies}. The intensity and the cosmic evolution of the EBL contain key details regarding the history of star formation and galaxy evolution \citep[e.g.,][]{madau2014cosmic}. Moreover, a proper characterization of the EBL is crucial for the accurate description of intrinsic properties of blazars \citep[e.g.,][]{dominguez2015spectral} {and the study of their earliest populations \citep[e.g.,][]{paliya20}}, the understanding of $\gamma$-ray propagation across cosmological scales \citep[e.g.,][]{aharonian1993very,biteau2022gamma,abdalla24,greaux24}, the detectability predictions of blazars at TeV energies \citep[e.g.,][]{paiano2021predictions,lainez25}, as well as proper estimation of cosmological {parameters \citep[e.g.,][]{blanch2005,barrau2008,fairbairn2013,biteau2015extragalactic,dominguez2019new,dominguez24a}. A precise measurement of the EBL intensity is also fundamental to the measurement of intergalactic magnetic field {\citep[e.g.,][]{dermer2011time,finke2015constraints,ackermann2018search}}, as well as in the quest of exotic physics such as the existence of axion-like particles {\citep[e.g.,][]{hauser1998cobe,kifune1999,kluzniak1999,protheroe2000,Amelino-Camelia2001,dominguez2011axion,buehler20}} and Lorentz invariance violation {\citep[e.g.,][]{abdalla2018lorentz,finke2023}}. It can also be instrumental in constraining the redshift of blazars \citep[e.g.,][]{prandini10,abdalla2020very,dominguez2024constraints}. \par 

 The direct measurement of the EBL intensity has proven challenging due to bright foreground contamination by the Zodiacal Light and the Diffuse Galactic Light \citep[e.g.,][]{hauser1998cobe,matsuura2011detection}. Some progress has been made in the near- and mid-infrared domain (NIR and MIR, respectively) using instruments onboard the \textit{Cosmic Background Explorer} \citep{hauser1998cobe} and subsequently, the \textit{Infrared Telescope in Space} \citep{matsumoto2005infrared}. {More recent results are reported by \citet{hsu24} using deep SCUBA-2 observations at 450 $\mu$m.} In the optical domain, the latest measurement has been made using the \textit{New Horizons} mission {\citep{lauer2021new,lauer2022,symons2023,postman2024new}.  This mission has made measurements of the EBL beyond the orbit of Jupiter, and thus with minimal contamination from the zodiacal foreground.} {Other really detailed analysis is carried out in a series of papers by the Hubble Space Telescope Archival Legacy program SKYSURF \citep[e.g.,][]{windhorst22}.} 
 \par 

An alternative approach is based upon the integration of galaxy counts, which allows a lower limit determination of the local, $z=0$, EBL \citep[e.g.,][]{dole2006cosmic}. The accuracy of this methodology is subject to the sensitivity of the survey and cosmic variance \citep{somerville2004cosmic}, since emission from the faint undetected galaxy populations, as well as from the outer regions of normal galaxies, may remain undetected. This is why these measurements are interpreted as lower limits on the EBL intensity {\citep[e.g.,][]{bernstein2002first,mattila2003}}. An updated measurement is presented by \citet{driver2016measurements}, who derived galaxy counts across 22 bands, ranging from the ultraviolet (UV) to the far-infrared (FIR). Their analysis incorporates data from several galaxy surveys, including the Cosmic Evolution Survey \citep[COSMOS;][]{scoville2007}, the Galaxy and Mass Assembly survey \citep[GAMA;][]{driver2011galaxy,koushan2021}, and the Cosmic Assembly Near-Infrared Deep Extragalactic Legacy Survey \citep[CANDELS;][]{grogin2011candels}. Subsequently, combining deep and extensive galaxy number counts from GAMA and Deep Extragalactic VIsible Legacy Survey \citep[DEVILS;][]{davies2018deep}, supplemented by the Hubble Space Telescope (HST) archive and other deep field observations, \cite{koushan2021} obtained EBL estimates showing good consistency with recent optical-EBL measurements derived from very high-energy (VHE) $\gamma$-ray observations. Their results match well with other independent methods (as discussed later), thereby placing more stringent constraints on the need for additional diffuse light beyond that contributed by the resolved galaxy population.

%\AD{[I think this part is outdated and should be changed by the new Driver's group paper on the GAMA/DEVILS survey, as detailed by Koushan et al. in 2021.]}
%Subsequently, evolving EBL intensity was determined using galaxy survey data involving multiwavelength observations of more than 150000 galaxies on the five different fields of CANDELS, containing a large number of galaxies till $z\sim6$ \citep{saldana2021observational}.  

On the theoretical side, efforts have been made to model the EBL and its cosmic evolution on the basis of different, complementary strategies. The models can be broadly classified into four different categories: (1) Forward evolution models employing semi-analytical models of galaxy formation \citep[e.g.,][]{gilmore2012semi,inoue2013extragalactic}, (2) backward evolution models which rely upon low redshift galaxy counts and are extrapolated to larger redshifts on the basis of some galaxy evolution model \citep[e.g.,][]{franceschini2008extragalactic,franceschini2017extragalactic,malkan2021gamma}, (3) another class of models converting the cosmic star formation history (SFH) to galaxy emission using some physical prescriptions \citep[e.g.,][]{finke2010modeling,khaire2015star,khaire19,finke2022modeling}, (4) finally, models based on some combination of galaxy luminosity density and galaxy spectra over a broad range of redshift \citep[e.g.,][]{dominguez2011extragalactic, scully2014empirical,saldana2021observational,koushan2021}. These models typically converge to spectral intensities that are comparable to, or even identical to, those inferred from galaxy counts, especially around the peak at $\sim\mu$m. {However, uncertainties increase near the FIR peak, mainly due to the scarcity of observational data at these wavelengths in most models and the difficulty of modeling FIR luminosity evolution, which is faster and more complex. Some efforts have been made to reduce these uncertainties \citep{saldana2021observational,finke2022modeling}.}

An indirect and powerful approach to constrain the EBL intensity is based on the detection of the absorption that the EBL imprints in the spectra of distant gamma-ray sources. Indeed, gamma rays traveling from large cosmological distances can interact, via the Breit-Wheeler pair-production mechanism, with  EBL photons, generating electron-positron pairs \citep{nikishov1962absorption,gould67EBL,fazio1970predicted}. This produces an attenuation in the spectra of distant gamma-ray sources, which can be used to constrain the intensity of the EBL.
The significant advancements in the sensitivity of gamma-ray telescopes, as well as in the modeling of the SEDs of blazars, have enabled many recent successes using this technique. 
In 2006, H.E.S.S. observations of two blazars, namely H 2356-309 and  1ES 1101-232, revealed the  EBL intensity to be very close to the lower limit obtained from galaxy counts \citep{aharonian2006low}. This was confirmed by subsequent observations of blazars at TeV and GeV energies \citep[e.g.,][]{aharonian2007new,magic2008very,finke2009constraints,abdo10constraints,georganopoulos2010method}. A breakthrough happened when large samples of blazars were used to measure the collective attenuation that the EBL is imprinting in their spectra \citep{ackermann2012imprint,abramowski2013measurement,biteau2015extragalactic,Abdollahi2018,desai2019,greaux24} and using that attenuation to deconvolve the intensity of the EBL across cosmic time \citep{finke2022modeling}. 

This work builds on the analysis of \cite{Abdollahi2018}, which used 101 months (Aug. 2008 to Jan. 2017) of {\it Fermi}-LAT \citep{atwood2009large} data, and 759 blazars in the redshift range $z=$0.03-3.1 to produce a measurement of the EBL optical depth across 12 epochs. Here, we update this measurement using a factor $\sim1.7$ larger exposure and an almost twice larger blazar sample (1464 vs 759 blazars) reaching  $z$=4.3 instead of $z$=3.1. This paper is organized as follows: $\S$~\ref{sec:sample} describes the sample and analysis technique, $\S$~\ref{sec:results} presents the measurements of the EBL optical depth, $\S$~\ref{sec:eblmodeling} derives the EBL intensity and its evolution with cosmic time, while $\S$~\ref{sec:discussion} discusses the results.
Throughout this paper we use a flat cosmology with $\Omega_M=0.3=1-\Omega_{\Lambda}$
and $H_0$=70 km s$^{-1}$ Mpc$^{-1}$.

\section{Sample and Analysis}
\label{sec:sample}
\subsection{Sample selection and analysis technique}
Our sample is selected from the objects reported in the fourth catalog of active galactic nuclei detected by the LAT, namely the 4LAC-DR3 catalog \citep{4lacAjello}. Although the 4LAC-DR3 includes a wider range of AGN subclasses, we
restrict our sample to blazars, since only relativistically beamed jet sources provide sufficiently bright and hard $\gamma$-ray spectra
extending to $\gtrsim 10$--100\, GeV where EBL attenuation becomes measurable. Other AGN types, such as radio-quiet Seyferts, Narrow Line Seyferts, and misaligned AGNs,  are typically fainter than blazars, making
them unsuitable for EBL absorption studies \citep{4lacAjello}. For the purpose of our analysis, all the blazars (519) lacking a redshift measurement are excluded. Moreover, we only retain blazars that are significantly detected in our analysis above 1 GeV (see below). Our final sample includes 752 FSRQs and 822 BL Lacs distributed between a redshift of 0.03 and 4.3. In Figure \ref{fig:red}, we show the redshift distribution of the blazars used in the study.  %\AD{A figure showing the redshift distribution of our sample in BL Lacs vs. FSRQs can be helpful for guiding the reader. Extra point: It may also be interesting to see if BL Lacs are now contributing to redshifts that were mostly constrained by FSRQs in the 2018 work.}
\begin{figure}
    \centering
    \includegraphics[width=0.7\linewidth]{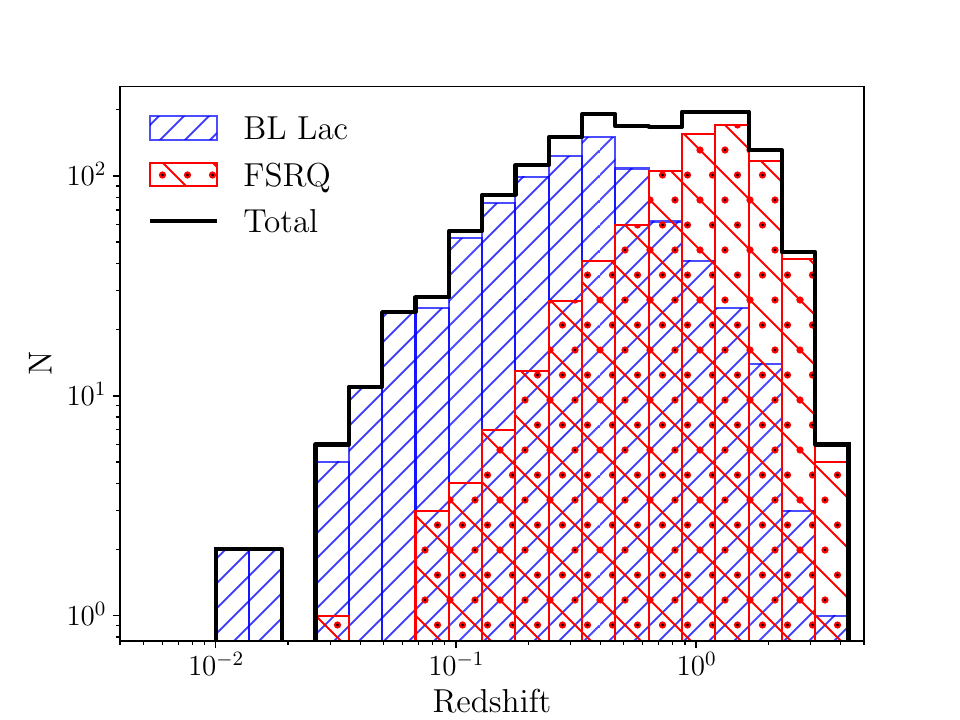}
    \caption{Redshift distribution of the sources used in this analysis on a logarithmic scale.}
    \label{fig:red}
\end{figure}
%\JB{This sounds pretty interesting. Maybe we could show a histogram of $\mathrm{TS_{EBL}}$ with redshift on the x-axis. Moreover, for example, we can make a side-by-side plot, the left panel could just be a simple histogram showing the redshift distribution of FSRQs and BL Lacs, and the right panel could show the histogram of $\mathrm{TS_{EBL}}$ values.} \par 

In the following, we adopt the same data-analysis methodologies as that of \cite{Abdollahi2018}.
Our analysis utilizes 181 months (Aug. 2008 to Sep. 2023) of Pass 8 \citep[P843;][]{bruel2018fermi} class `SOURCE' photons detected by the LAT between 1 GeV and 1 TeV. The times when the spacecraft was over the South Atlantic Anomaly are not considered, and the photons detected
at angles larger than $100^\circ$ relative to the zenith are removed. For the analysis of each source, we consider a square region of interest of size $10^\circ \times 10^\circ$
 centered on the source position (region of interest, ROI). For the analysis of each ROI, we define a sky model comprising the latest isotropic emission model ($\text{iso\_P8R3\_SOURCE\_V3\_v1.txt}$\footnote{\url{https://fermi.gsfc.nasa.gov/ssc/data/access/lat/BackgroundModels.html}}) and the latest diffuse Galactic emission model ($\text{gll\_iem\_v07.fits}$\footnote{\url{https://fermi.gsfc.nasa.gov/ssc/data/access/lat/BackgroundModels.html}}).  We use the standard data filters (DATA QUAL $>0$  and LAT CONFIG==1) and use the $\text{P8R3\_SOURCE\_V3}$ IRFs. We begin the analysis by optimizing the model components for the ROI of each target using a binned maximum-likelihood fit, and evaluate the significance of each source in the ROI using the TS defined as:
\begin{equation}
    TS = 2\log{\frac{\mathcal{L}}{\mathcal{L}_0}},
\end{equation}
where $\mathcal{L}_0$ denotes the likelihood for the null hypothesis (i.e., there is no source, just background) and $\mathcal{L}$ is the likelihood for the alternative hypothesis (there is a source plus background). The spectral parameters of the Galactic diffuse component (index and normalization) and the normalization of the isotropic component are left free to vary. The normalizations for all the 4FGL sources with $TS > 25$ within $5^\circ$ of the ROI center, as well as sources with $TS > 500$ and within $7^\circ$ of the ROI center, are also kept free. We also use the {\tt Fermipy} function \textit{find\_sources} in order to search for any unmodeled residuals, i.e. potential point sources not included in the catalog. This function generates TS maps and finds new sources on the basis of the peaks in the TS map. The maps are generated using a power-law spectral model with a power-law index of 2. The minimum
separation between two point sources is set to be $0.5^\circ$, and the minimum TS for including a source in the model is set to 16. Any new source is included in the ROI model with optimized, best-fit, parameters.
 At the end of this stage we evaluate the TS of the target source (the blazar of interest) at the center of the ROI and only retain those blazars that are significantly detected with  $TS \geq 25$.

\subsection{Intrinsic spectra of blazars}
We model the spectrum of each blazar as an intrinsic spectrum attenuated by the EBL. Following \cite{ackermann2012imprint} and \cite{Abdollahi2018}, we allow for the optical depth provided by a given model to be renormalized by a factor $b$, which is fit to the data of all blazars in a joint-likelihood fit.
The observed and intrinsic spectra of a given blazar are related as follows:
\begin{equation}
    \left(\frac{dN}{dE}\right)_{obs} = \left(\frac{dN}{dE}\right)_{int}e^{-b\tau_{model}(E,z)}, 
    \label{eq:intrinsic}
\end{equation}
where $\left(\frac{dN}{dE}\right)_{int}$ and $\left(\frac{dN}{dE}\right)_{obs}$ are the intrinsic and observed spectra, $\tau_{model}(E,z)$ is the optical depth as a function of energy, at the source redshift, as estimated by the chosen EBL model, while $b$ is the renormalization factor.

In order to model the intrinsic spectra of blazars, we follow the strategy of \cite{Abdollahi2018}, which was optimized on simulations. In short, the model is fitted to the data only to a maximum energy ($E_{max}$) up to which the attenuation of the EBL is negligible. This is defined
as the energy at which the optical depth for the model of 
\cite{finke2022modeling} is $\tau_{\gamma\gamma} < 0.1$ (see Fig.~\ref{fig:emax}). We have tested using several EBL models to define $E_{\rm max}$ and found it had little impact on our results.

\begin{figure}
    \centering
    \includegraphics[width=0.5\linewidth]{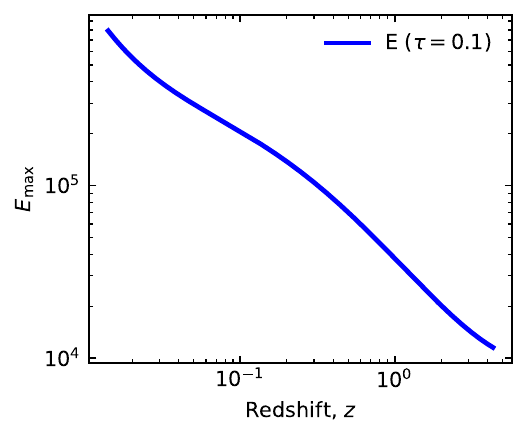}
    \caption{\(E_{\max}\) versus redshift for the blazar sample. \(E_{\max}\) corresponds to the energy at which the optical depth satisfies \(\tau_{\gamma\gamma} = 0.1\) for the EBL model of \cite{finke2022modeling}.}

    \label{fig:emax}
\end{figure}

Then, we adopt the log-parabola as our baseline spectral model. This model is defined as:
\begin{equation}
    \left( \frac{dN}{dE} \right)_{int} = N_0\left(\frac{E}{E_b}\right)^{-\alpha+\beta\log(E/E_b)}, 
\end{equation}
where $N_0$ (normalization), $\alpha$ (photon index) and $\beta$ (curvature) are the free parameters and $E_b$ is the scale factor fixed at 1 GeV. 
We then test whether a power-law model with an exponential cutoff provides a better representation of the data. This model is defined as:
\begin{equation}
    \left( \frac{dN}{dE} \right)_{int} = N_0\left(\frac{E}{E_b}\right)^\alpha e^{-(E/E_b)^{\gamma_1}}, 
\end{equation}
where $E_c$ (the cutoff energy) and $\gamma_1$ (the exponential index) are free parameters. %Broken power-law and smoothed broken power-law were never found to choose the blazer spectra better than the above-mentioned models. \par 

Following \cite{Abdollahi2018}, we define two TS of curvature $TS_{c,1}$ and $TS_{c,2}$ in the following way:
\begin{equation}
    TS_{c,1} = 2(\log{L_{exp,\gamma_1=0.5}} - \log{L_{log-parabola}}),
\end{equation}
\begin{equation}
    TS_{c,2} = 2(\log{L_{exp,\gamma_1=free}} - \log{L_{log-parabola}}). 
\end{equation}
where $\log{L_{exp,\gamma_1=0.5}}$ and $\log{L_{exp,\gamma_1=free}}$ are the log- likelihoods derived using the exponential cut-off model with $\gamma_1 = 0.5$ \citep[supported by the observation of hundreds of {\it Fermi}-LAT FSRQs, see][]{ajello2012} and $\gamma_1$ being kept free to vary, respectively, while $\log{L_{log-parabola}}$ is the log-
likelihood of the log-parabola model.  We use the thresholds used in \cite{Abdollahi2018} and reported in Table~\ref{tab:thresholds}  to choose the model to describe the spectrum of a given blazar.
\begin{center}
%\caption{Table}
%\begin{tabular}{|c c c |} 
\begin{deluxetable*}{lccccc}
\tablecaption{Thresholds for selecting blazars' intrinsic models \label{tab:thresholds}}
\tablehead{ $TS_{c,1}$ & $TS_{c,2}$ & Model chosen}
 \startdata
 $<1$ & $<3$ & Log-parabola \\ 
$>1$ & $<3$ & Power law with cutoff $\gamma_1=0.5$ \\
$>1$ & $>3$ & Power law with cutoff with $\gamma_1$ free \\
%\end{tabular}
\enddata
\end{deluxetable*}
\end{center}

%In the analysis procedure, $\gamma_1$ remains fixed at 0.5 or at the best-fitted value in order to avoid the convergence problem.  \par 

%The spectral parameters for all sources barring the source of interest are frozen at their best fit value. Subsequently, we apply the EBL attenuation on the spectrum of the target source. Based on the above mentioned criteria, we update the source spectrum either to \textit{EBLAtten::PowerLaw2} or \textit{EBLAtten::LogParabola}. 

\begin{figure*}
\gridline{\fig{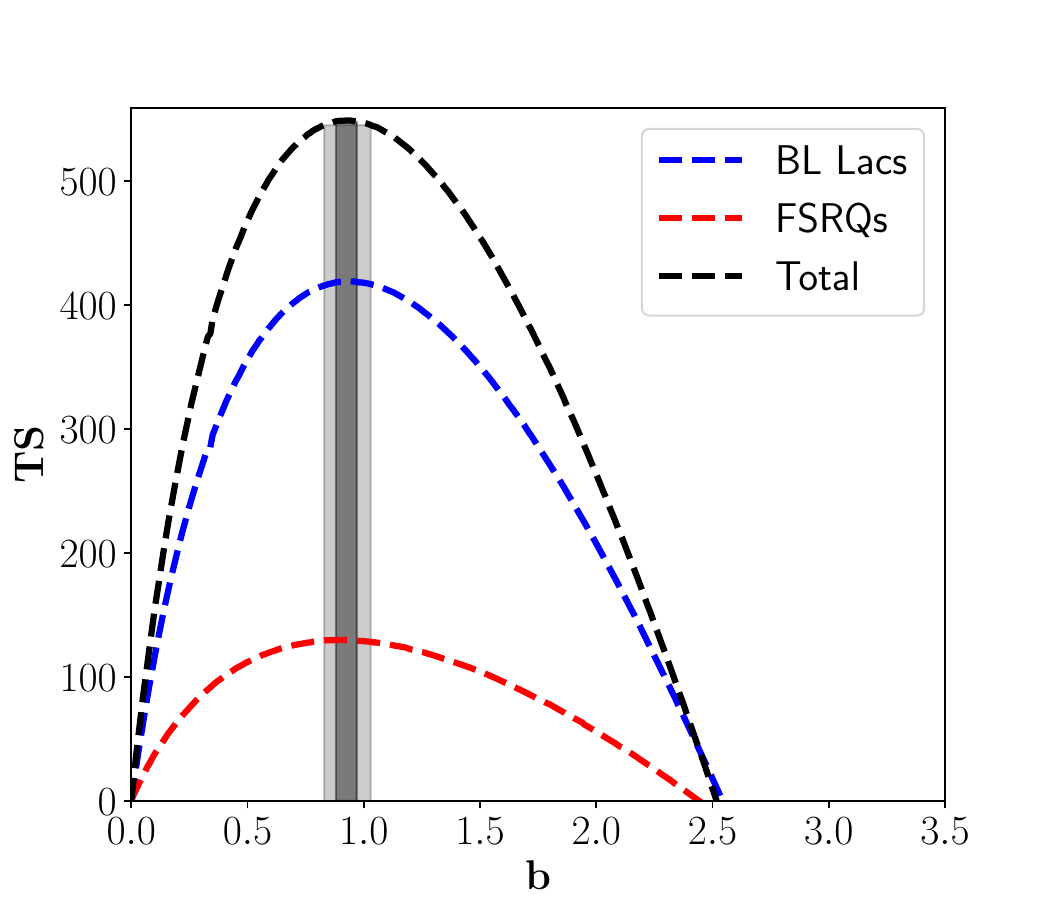}{0.35\textwidth}{(a)}
          \fig{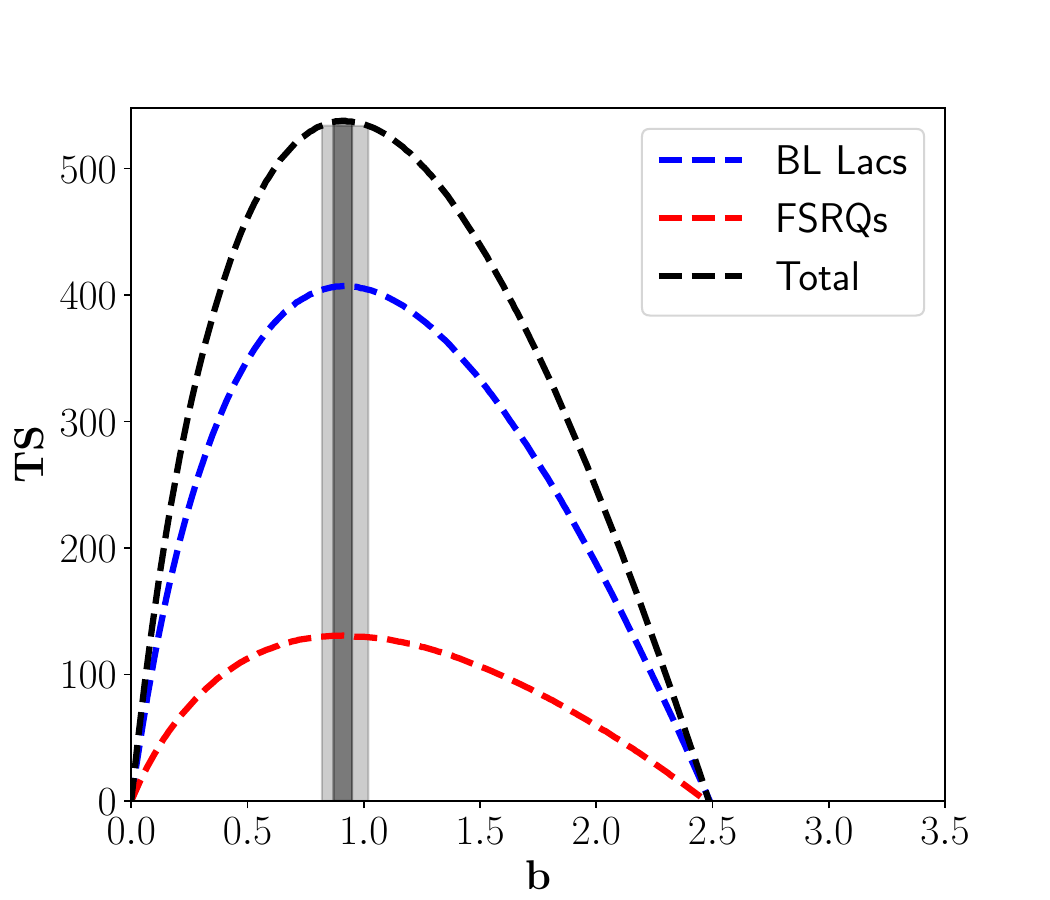}{0.35\textwidth}{(b)}
          \fig{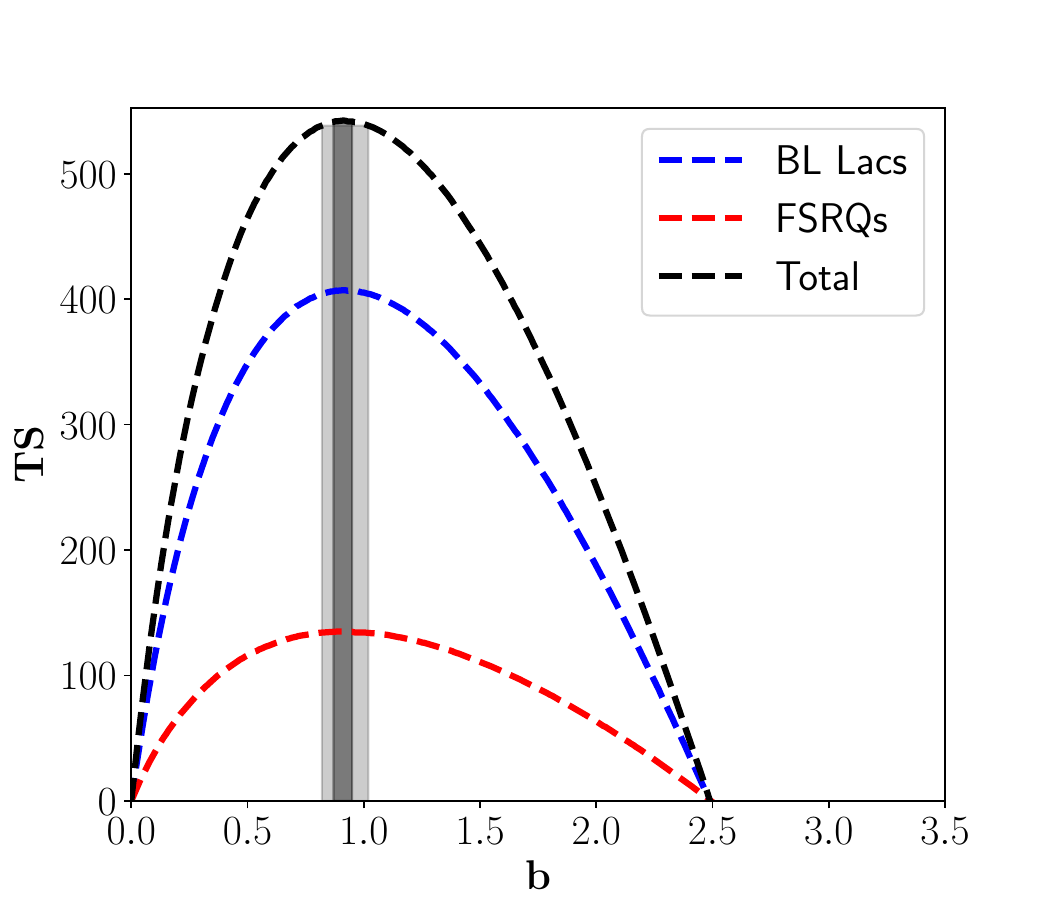}{0.35\textwidth}{(c)}
          }
\caption{TS$_{\mathrm{EBL}}$ profiles of the EBL detection as a function of the renormalizing parameter $b$, for the EBL models of \citet[][left]{finke2022modeling}, \citet[][center]{saldana2021observational} and \citet[][right]{franceschini2008extragalactic}.
The dark and light shaded regions show the $1\sigma$ and $2\sigma$ confidence intervals around the best-fit $b$ values. The dashed red, blue and black lines correspond to the FSRQ, BL Lac and combined $TS$ profiles, respectively.  The best-fit parameters are reported in Table~\ref{tab:models}.
%The best-fit $b$ corresponds to $b=0.93\pm0.10$, with the $TS$ profile peaking at $TS = 543$. (b) Same as (a), for the model \citep{saldana2021observational}. The best-fit $b$ corresponds to $b=1.25\pm0.10$, with the $TS$ profile peaking at $TS = 553$. (c) Same as (a), for the model \citep{franceschini2017extragalactic}. The best-fit $b$ corresponds to $b=1.04\pm0.10$, with the $TS$ profile peaking at $TS = 471$.
\label{fig:stacked}}
\end{figure*}

%\AD{All figures needs to go in high-resolution format and not png. The x-axis can go up to 3.5, there is no info above that value, increase font size by a lot.}

%\AD{Are we say in the intrinsic spectra section, how the uncertainties in the intrinsic spectra are propagated? or, at least, mention something in this line?}

\begin{deluxetable*}{lcc}
\tablecaption{Joint-Likelihood fits for different EBL models \label{tab:models}}
\tablehead{Model & TS$_{\mathrm{EBL}}$ & $b$}
    \startdata
    \cite{finke2022modeling} & 543 &  $0.93\pm0.10$\\
    \cite{saldana2021observational} & 538 & $0.91\pm0.11$\\
    \cite{franceschini2017extragalactic} & 542 & $0.91\pm0.11$
    \enddata
\end{deluxetable*}

%%%%%%%%%%%%%%%%%%%%%%%%%%%%%%%%%%%%%%%%%%%%%%%%%%%%%%%%%%%%%%%%%%%%
%
%
%    RESULTS
% 
%
%%%%%%%%%%%%%%%%%%%%%%%%%%%%%%%%%%%%%%%%%%%%%%%%%%%%%%%%%%%%%%%%%%%%%

\section{Results}
\label{sec:results}
\subsection{Joint-Likelihood fit}\label{stacking}
\label{sec:ebl}
Once the intrinsic spectra of blazars have been determined, we can establish how significantly the EBL attenuation is detected in all of them. We do this by reverting the energy to 1\,GeV - 1\,TeV and modeling each source spectrum as Eq.~\ref{eq:intrinsic}.
All the ROIs and blazars have only the $b$ parameter in common. An EBL attenuation as strong as estimated by the adopted EBL model would imply $b=1$, while $b=0$ implies the absence of EBL attenuation. We define $TS_{EBL}$ as:
\begin{equation}
TS_{EBL}=2log\frac{\mathcal{L}(b)}{\mathcal{L}(b=0)},
\end{equation}
where $\mathcal{L}(b)$ is the log-likelihood obtained when $b$ is fit to the data and $\mathcal{L}(b=0)$ is the log-likelihood of the case where there is no EBL. The $TS_{\mathrm{EBL}}$ measures the statistical significance with which the hypothesis of no EBL attenuation in the spectra of blazars can be rejected. For each {blazar}, we scan the $TS_{\mathrm{EBL}}$ by varying $b$ in small intervals. We then sum all the $TS_{\mathrm{EBL}}$ profile for all blazars together. We evaluate the $TS_{\mathrm{EBL}}$ using three of the newest EBL models \citep{franceschini2017extragalactic,saldana2021observational,finke2022modeling} and report the results in Figure~\ref{fig:stacked}, while {Table}~\ref{tab:models} summarizes the results. For all three models, the $TS_{\mathrm{EBL}}$ between 538 and 543, which implies a collective detection of the EBL attenuation with a significance in the $\sim23$\,$\sigma$ range. We find that the $TS_{\mathrm{EBL}}$ has increased by $\approx$80\,\% with respect to our previous analysis reported in \cite{Abdollahi2018}. This is due to the larger exposure and larger source sample, than \cite{Abdollahi2018}, used in this work. The strength of the EBL attenuation observed in the {\it Fermi}-LAT spectra is in good agreement, with the estimates of all the models, as testified by the $b$ parameters being consistent with 1.0 within 2$\sigma$ uncertainties. %Indeed, inspection of {Figure~12 from \cite{saldana2021observational} shows that their} model predicts a lower EBL intensity than that measured by {\it Fermi}-LAT in \cite{Abdollahi2018} at 1\,$\mu$m across the $0.8<z<2.0$ range.

\begin{figure}
    \centering
    \includegraphics[width=0.7\linewidth]{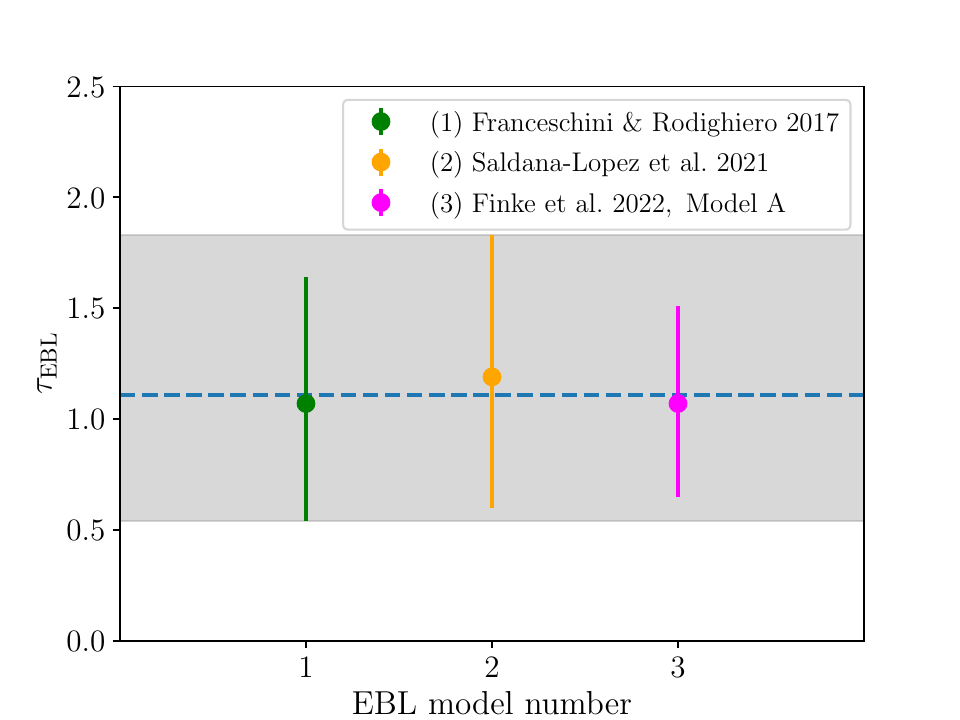}
    \caption{Illustration of the computation of the optical depth for the  $0.70\leq z<0.79$ bin and photon energies between 215 GeV - 464 GeV using the $\tau$-values corresponding to the three different models. The horizontal dashed line is the average optical depth derived using the three models, while the gray band shows the uncertainty encompassing the uncertainty of all models.}%\JF{[The legend for the figure has a few typos in it.  The first model should be Saldana-Lopez (no spaces before or after dash) {\bf\em et} (not el) al. 2021.  Please capitalize ``model'' in the Finke et al.\ 2022 line.]}\AD{Sort the references in the legend by date from early to late, actually a better and nicer solution is to change the numbers in the x-axis for the rotated, maybe 45 deg, model names, it's easy to do}}
    \label{fig:alltau}
\end{figure}

\begin{figure}
    \centering
    \includegraphics[width=0.98\linewidth]{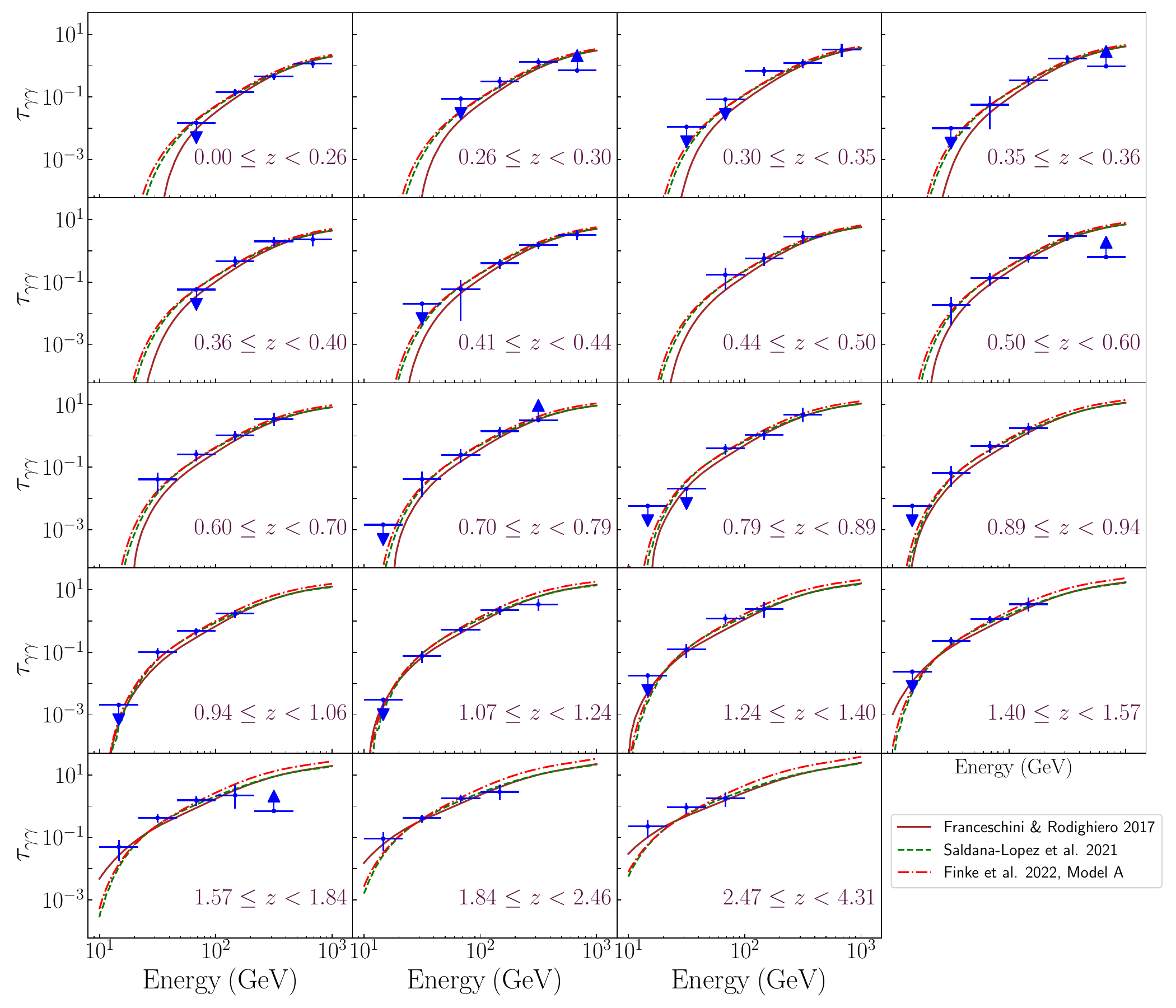}
    \caption{Measurements of the EBL optical depth, $\tau_{\gamma\gamma}$, as a function of energy and redshift. The solid lines show the estimates of three different EBL models.}
    %\JF{[typo in legend.  Saldana-Lopez {\bf\em et} al.  As it is now set up, I think the y-axis should just be $\tau_{\g\g}$, not $\log\tau_{\g\g}$.  ]}
    %\AD{It will be nicer to keep the same color coding for the models in all figures, are the long uncertainties at the bottom panels real or they should be upper limits? please double check}}
    \label{fig:tau}
\end{figure}

\subsection{Measurements of the Optical Depth and Its Uncertainty}
\label{sec:tau}
To measure the EBL optical depth as a function of redshift and energy, we repeat the same process above ($\S\ref{sec:ebl}$), but in smaller energy and redshift bins \citep[see also][]{chen2004,Abdollahi2018}.  
The redshift bins are chosen such that in each bins there is a sufficient number of sources to achieve a detection of the EBL at $TS_{\mathrm{EBL}}\geq$25. For each redshift bin, we chose 6 logarithmically-spaced energy bins between 1\,GeV and 1\,TeV.  The best-fit $b$ value in a given energy and redshift bin is converted to a measurement of $\tau_{\gamma\gamma}$ following the method adopted {by} \cite{Abdollahi2018} and \cite{desai2019}. For each of the three EBL models in {Table}~\ref{tab:models}, we compute $\tau_i(E,z)$ as {$\tau_i(E,z)=b\times\tau_{model,i}(E,z)$}.
We compute $\tau_{\gamma\gamma}$ as the average of the three models, with the associated uncertainty defined to encompass the uncertainties of all three.
{This brackets the uncertainty in the overall shape of the EBL absorption.} Figure~\ref{fig:alltau} shows a visual representation of how the $\tau_{\gamma\gamma}$ is derived. For upper and lower limits, we choose the most conservative limit provided by any of the models. In Table \ref{tab:tau}, we report the optical depths in different redshift and energy intervals, as well as the weighted redshift in each redshift interval ($z_{\rm eff}$). The effective redshift for each bin was calculated as the TS-weighted mean, i.e., by summing $z_i \times \mathrm{TS}_i$ for all sources in the bin and dividing by the total TS in that bin.

The systematic uncertainties of this analysis are detailed in \cite{Abdollahi2018} and are related to the choice of maximum energy ($E_{max}$) used to determine the intrinsic spectrum, uncertainties in the intrinsic spectrum, and uncertainties related to the instrumental response function (IRF). The uncertainty related to the maximum energy used to determine the intrinsic spectrum can be assessed by adopting an $E_{max}$ energy of 10\,GeV for all the spectra, independently of the source redshift. When doing so, we observe variation in the best-fit $b$ values of up to 10\,\%. Relaxing the conditions in Table~\ref{tab:thresholds} allows a way to assess the uncertainty in the determination of the intrinsic spectra. We observe a variation in the best-fit $b$ values of up to 13\,\% if we fit all sources with a Log-parabola model or if we only allow the source spectra to be modeled with a power law with exponential cut-off if $TS_{c,2}>3$. Finally, uncertainties in the IRF introduce a 10\,\% uncertainty on $b$ as detailed in \cite{ackermann2012imprint} and \cite{Abdollahi2018}.  
To account for the uncertainty above, we sum in quadrature $0.2\times \tau_{\gamma\gamma}$ to the statistical uncertainty in each bin.

An additional uncertainty could be related to BL Lacs with inaccurate redshift measurements. It is not uncommon to find different redshifts for the same BL Lac in the literature. 
In order to understand the impact of inaccurate redshifts, we performed tests where we assigned a random redshift (extracted from the distribution of BL Lacs in Fig.~\ref{fig:red}) to up 30\,\% of the BL Lacs in our sample.
We find that if 10\,\% of the BL Lacs in our sample have a wrong redshift, the results remain compatible with our results, within the systematic uncertainty reported above. For larger fractions, the results are not compatible. For example, if we extract random redshifts for 30\,\% of the BL Lac sample, we obtain $b=0.65\pm0.10$ and $TS_{EBL}\approx170$. Moreover, the results obtained from the FSRQ sample and the BL Lac one become incompatible. In summary, our tests rule out that a large fraction of BL Lacs have an inaccurate redshift and show that our analysis is robust against small fractions ($\leq10$\,\%) of BL Lacs  having an inaccurate redshift.

Figure \ref{fig:tau} shows the measurements of the optical depth in 19 bins of redshift and 6 energy bins. Because of the larger exposure and larger source sample, this analysis {expands} the measurement of the EBL optical depth from 12 to 19 epochs, {and reaches a higher} redshift. In every redshift bin, our measurement is most sensitive around $\tau_{\gamma\gamma}=1$ the transition between a transparent to an opaque Universe. Below that, the EBL attenuation is too small to be detected, while for large $\tau_{\gamma\gamma}$, {too few gamma rays survive for the absorption to be measurable}. %As anticipated in $\S$~\ref{sec:ebl}, the \cite{saldana2021observational} model predicts a lower EBL optical depth in the $0.3<z<0.8$ range than measured and than predicted by the \cite{franceschini2008extragalactic} and \cite{finke2022modeling} models.

%We have measured the optical depth as a function of redshift and energy by renormalizing the model-predicted optical depth in small redshift and energy bins. The redshift bins were selected such that each bin contains roughly equal signal strength, i.e. similar values of $TS_{EBL}$. The energy bins were selected to ensure that they have equal logarithmic widths. Corresponding to each redshift and energy bin, a stacked $TS_{EBL}$ vs $b$ profile is generated following the same procedure illustrated in Section \ref{stacking}, where the source sample is modified as per the redshift bin. The optimized $b$ value as well as the EBL model under consideration are then used to obtain the optical depth $\tau_{\gamma\gamma}$ due to EBL attenuation. This analysis is performed for three different EBL models \citep{finke2022modeling,franceschini2017extragalactic,saldana2021observational}. Final optical depth in a given redshift and energy bin is the mean of the optical depth values obtained from all the three models, while the uncertainties are considered to be the one encompassing all the optical depth measurements, as demonstrated in Figure \ref{fig:alltau}. In Figure \ref{fig:tau} we show the measurement of optical depth $\tau_{\gamma\gamma}$ in different redshift and energy bins. 

\begin{deluxetable*}{cccccccc}
%\tablenum{2}
%\tablecaption{ {Optical depth \label{tab:tau}\label{tab:tau}}

\tablecaption{Optical depth values in different redshifts and energy intervals. \label{tab:tau} }%\JF{[Would it make sense to add another column, with the $z_{\rm eff}$, or something like that?  I think people will want to know what one redshift they can use for each energy bin.]}\AD{[The energies of the bins should be given as an integer, without the decimal point. The optical depths of the limits should be given with two decimal points as the rest of the values and not with three decimal points.]}}
\tablewidth{0pt}
\tablehead{
\colhead{Redshift} & \colhead{$z_{\rm eff}$} & \colhead{10-21} & \colhead{21-46} & \colhead{46-100} & \colhead{100-215} & \colhead{215-464} & \colhead{464-1000} \\
\colhead{($z$)} &  & \colhead{(GeV)} & \colhead{(GeV)} & \colhead{(GeV)} & \colhead{(GeV)} & \colhead{(GeV)} & \colhead{(GeV)} \\
%\colhead{} & \colhead{} & \colhead{(2024 May 27)} & \colhead{(2024 July 5)}\\
}
%\decimalcolnumbers
\startdata
%diskbb & $kT_{in}$ & $0.73\pm0.38$ & --   \\
%diskbb & norm & $0.97\pm0.23$ & --   \\
%\hline
0.01-0.26 & 0.19 & - & - & $<0.01$ & $0.14_{-0.03}^{+0.03}$ & $0.45_{-0.06}^{+0.06}$ & $1.17_{-0.17}^{+0.19}$ \\
0.26-0.30 & 0.28 & - & - & $<0.02$ & $0.31_{-0.13}^{+0.14}$ & $1.32_{-0.33}^{+0.39}$ & $>1.11$ \\
0.30-0.35 & 0.32 & - & $<0.02$ & $<0.07$ & $0.67_{-0.16}^{+0.17}$ & $1.23_{-0.32}^{+0.36}$ & $3.23_{-1.17}^{+1.71}$ \\
0.35-0.36 & 0.35 & - & $<0.01$ & $0.05_{-0.04}^{+0.04}$ & $0.34_{-0.08}^{+0.09}$ & $1.68_{-0.30}^{+0.34}$ & $>1.01$ \\
0.36-0.40 & 0.38 & - & - & $<0.054$ & $0.46_{-0.14}^{+0.16}$ & $2.01_{-0.49}^{+0.60}$ & $2.31_{-0.81}^{+1.04}$ \\
0.41-0.44 & 0.42 & - & $<0.004$ & $0.06_{-0.05}^{+0.05}$ & $0.40_{-0.11}^{+0.11}$ & $1.53_{-0.30}^{+0.30}$ & $3.28_{-0.89}^{+1.12}$ \\
0.44-0.50 & 0.47 & - & - & $0.17_{-0.10}^{+0.11} $ & $0.57_{-0.22}^{+0.26}$ & $2.83_{-0.94}^{+1.33}$ & $ - $ \\
0.50-0.60 & 0.57 & - & $0.02_{-0.01}^{+0.01}$ & $0.13_{-0.05}^{+0.06}$ & $0.60_{-0.12}^{+0.14}$ & $2.99_{-0.67}^{+0.86}$ & $>0.64$ \\
0.60-0.70 & 0.66 & - & $0.04_{-0.02}^{+0.02}$ & $0.26_{-0.09}^{+0.10}$ & $1.04_{-0.27}^{+0.31}$ & $3.47_{-1.25}^{+1.99}$ & - \\
0.70-0.79 & 0.74 & $<0.001$ & $0.04_{-0.03}^{+0.03}$ & $0.25_{-0.10}^{+0.11}$ & $1.43_{-0.39}^{+0.44}$ & $>3.21$ & - \\
0.79-0.89 & 0.86 & $<0.005$ & $<0.017$ & $0.40_{-0.12}^{+0.13}$ & $1.09_{-0.30}^{+0.34}$ & $4.80_{-1.71}^{+3.07}$ & $ - $ \\
0.89-0.94 & 0.92 & $<0.005$ & $0.06_{-0.05}^{+0.05}$ & $0.53_{-0.19}^{+0.20}$ & $2.06_{-0.69}^{+0.87}$ & - & - \\
0.94-1.06 & 1.00 & $<0.002$ & $0.10_{-0.02}^{+0.02}$ & $0.47_{-0.10}^{+0.10}$ & $1.72_{-0.34}^{+0.41}$ & - & - \\
1.07-1.24 & 1.18 & $<0.003$ & $0.07_{-0.02}^{+0.02}$ & $0.52_{-0.07}^{+0.07}$ & $2.27_{-0.52}^{+0.65}$ & - & - \\
1.24-1.40 & 1.33 & $<0.01$ & $0.12_{-0.05}^{+0.05}$ & $1.20_{-0.29}^{+0.33}$ & $2.42_{-1.01}^{+1.50}$ & - & - \\
1.40-1.57 & 1.50 & $<0.02$ & $0.23_{-0.04}^{+0.04}$ & $1.14_{-0.29}^{+0.31}$ & $3.44_{-1.22}^{+1.98}$ & - & - \\
1.57-1.84 & 1.75 & $0.04_{-0.03}^{+0.03}$ & $0.42_{-0.10}^{+0.10}$ & $1.54_{-0.43}^{+0.51}$ & $2.21_{-1.31}^{+2.53}$ & $>0.70$ & - \\
1.84-2.46 & 2.15 & $0.09_{-0.05}^{+0.05}$ & $0.41_{-0.08}^{+0.09}$ & $1.76_{-0.39}^{+0.47}$ & $2.87_{-1.20}^{+1.94}$ & - & - \\
2.47-4.31 & 2.59 & $0.22_{-0.12}^{+0.13}$ & $0.90_{-0.26}^{+0.29}$ & $1.77_{-0.74}^{+0.96}$ & - & - & - \\
\hline
\enddata
\end{deluxetable*}

\subsection{Cosmic Gamma-ray Horizon}
\begin{figure}
    \centering
    \includegraphics[width=0.7\linewidth]{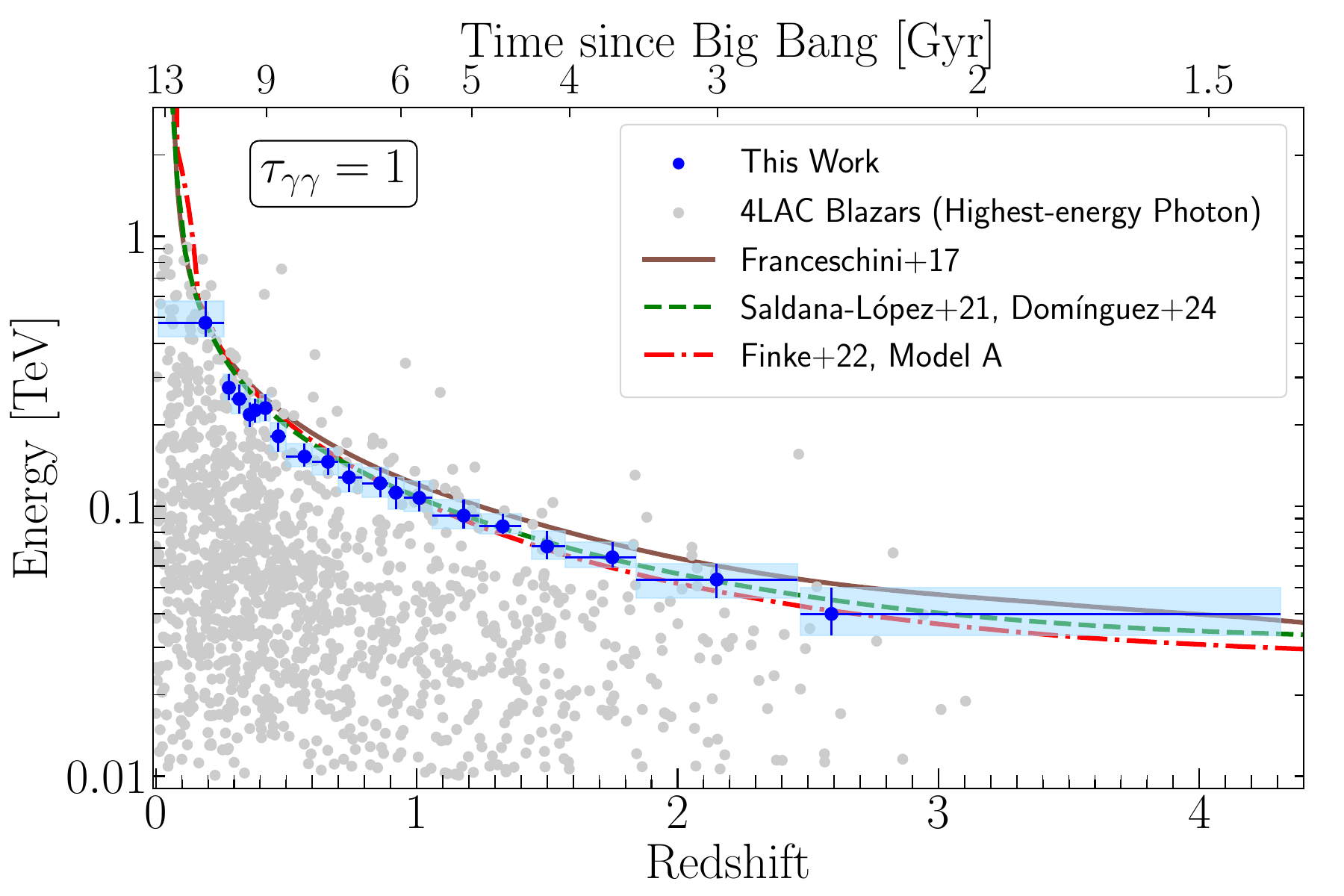}
    \caption{Measurement of the cosmic $\gamma$-ray horizon derived from the EBL optical depth measurements (blue data points). The lines correspond to the estimates of three EBL models (described in the legend). The gray data points are the single most energetic photon detected from each source used in this analysis.}
    \label{fig:cghr}
\end{figure}
The cosmic gamma-ray horizon (CGRH) is defined as the energy $E_{\mathrm{CGRH}}(z)$ at which the optical depth $\tau_{\gamma\gamma}(E, z) = 1$, marking the transition between the transparent and opaque Universe.
 This threshold is an integral measure of the evolving photon field, from UV to FIR, and encapsulates the cumulative star formation and dust reprocessing over cosmic history.
The CGRH is well detected in each redshift bin of Figure~\ref{fig:tau} and is plotted as function of redshift in Figure~\ref{fig:cghr}.

The updated CGRH is in agreement with the optical depths derived from the empirical EBL model of \citet{saldana2021observational}, as implemented in \citet{dominguez2024constraints}; the model by \citet{finke2022modeling}, which incorporates the cosmic star formation history and dust reprocessing; and the model of \citet{franceschini2017extragalactic}, which uses fits to a wide variety of multi-wavelength data. The Saldana-Lopez model benefits from deep multiwavelength galaxy observations, particularly from CANDELS, while the Finke model offers a complementary perspective based on physical prescriptions for galaxy evolution. The \citet{franceschini2017extragalactic} model is an update of the \citet{franceschini2008extragalactic} model with new {\em Herschel} and {\em Spitzer} data.  This model derives a slightly more transparent universe than the other two models, but is still broadly consistent with the upper error bars of the measured CGRH.
The consistency of our results with the models reinforces the robustness of the \textit{Fermi}-LAT gamma-ray attenuation method and supports the validity of current EBL modeling across a range of methodologies and assumptions.

%Compare with \citet{arsioli2025} .

\subsection{Time-Resolved Analysis}
Blazars exhibit intrinsic variability across all wavelengths, which can introduce biases and complexities in the measurement of the EBL attenuation. To address whether there is a bias, we conducted a time-resolved analysis on sources with the highest variability index and the largest $TS_{EBL}$. Two sources stand out among the rest: 4FGL J0538.8$-$4405 (PKS~0537$-$441, BL Lac, $z=0.894$, $TS_{EBL} = 10.2$), which has interesting multiwavelength variability and spectral properties \citep{dammando2013};  and 4FGL J1504.4+1029 (PKS~1502+106, FSRQ, $z=1.837$, $TS_{EBL} = 7.04$), {which has been suggested as a possible neutrino source \citep{rodrigues2021,blinov2025}}. 
We produced two-week binned lightcurve in the 1\,GeV-1\,TeV energy range, for the 181\,months of this analysis, for both sources and used the Bayesian Block (BB) analysis \citep{scargle1998} to define times when the source changed state. For each of the BBs, we extract the source spectrum and repeat the analysis in  $\S$\ref{sec:ebl} to derive the $TS_{EBL}$ vs $b$ profile. We then sum the profiles of all BBs together. 
We measured the following $b$ values for  4FGL J0538.8-4405  and 4FGL J1504.4+1029, respectively: 0.94$\pm0.44$ and 1.39$\pm0.75$. When compared to the time-averaged ones ($b=1.19\pm0.42$ and $b=1.01\pm0.41$), we find that they are consistent {within 1$\sigma$}, suggesting that variability has a negligible impact on the overall measurement of the EBL attenuation. In Figure~\ref{fig:bblock}, we present the light curve of 4FGL J1504.4+1029 along with its Bayesian block decomposition.

%Figure \ref{fig:bblock} presents the light curve of 4FGL J1504.4+1029 along with its Bayesian block decomposition. Within each block, the optimal intrinsic model is determined based on the criteria outlined in Table 1. Since time-resolved spectra for a given source can be treated as independent observations, their contributions to $TS_{EBL}$ are summed with the rest of the sample. The best-fit $b$-value remains consistent within the $1\sigma$ uncertainty range, suggesting that variability has only a marginal impact on the overall measurement of EBL attenuation.

\begin{figure}
    \centering
    \includegraphics[width=0.75\linewidth]{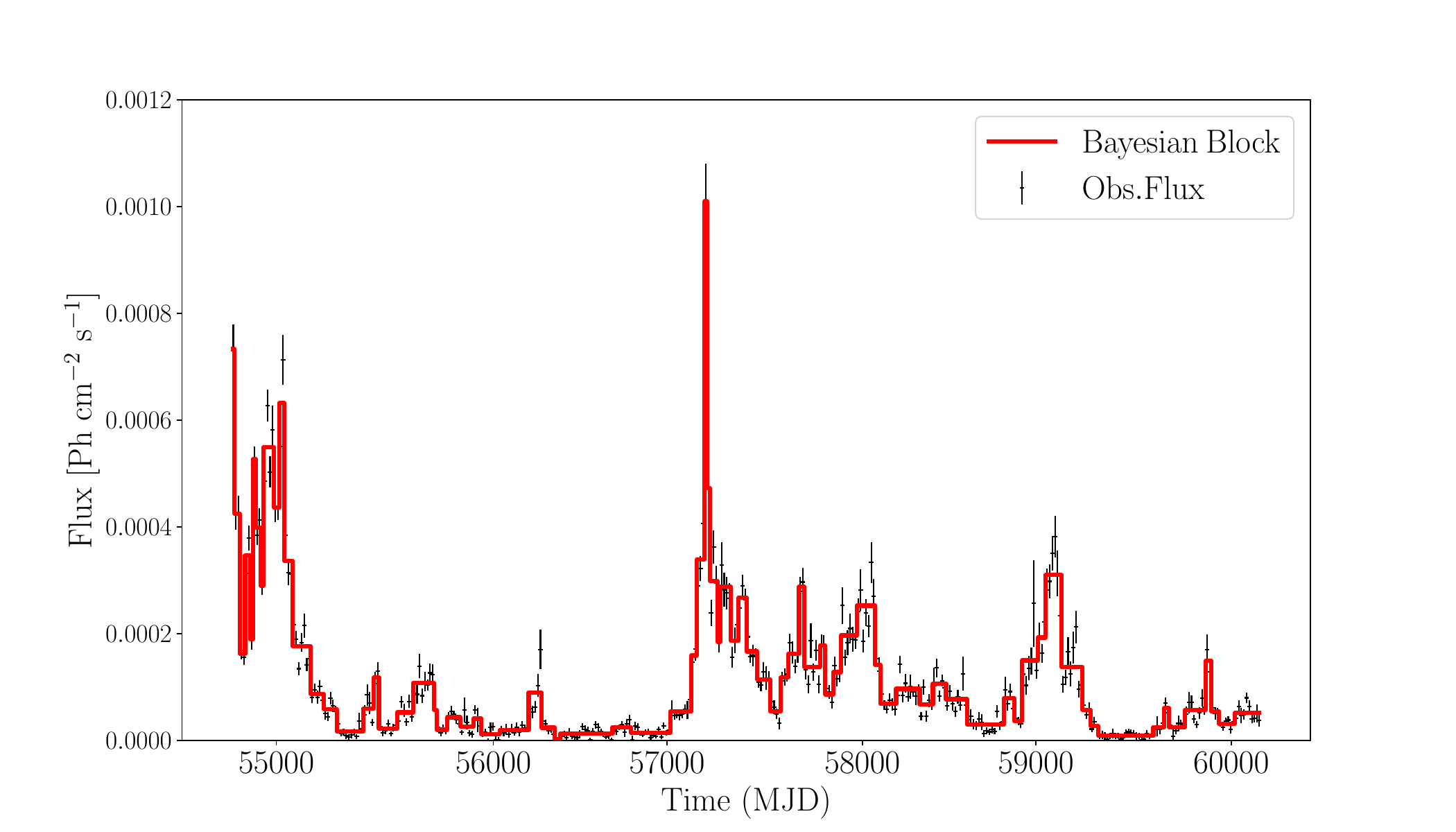}
    \caption{Lightcurve of 4FGL J1504.4+1029 produced using a two-week bin. The red line shows the Bayesian Blocks.} %\JF{[Could you make the MJD labels more round number intervals?  Like 55000, 56000, 57000, etc.  For the y-axis, maybe label it ``Flux [ph cm$^{-2}$\ s$^{-1}$]''.  Also I find the text for this figure kind of hard to read.  Is it possible to produce a higher resolution version, and make the text a bit bigger?]}}
    \label{fig:bblock}
\end{figure}

%%%%%%%%%%%%%%%%%%%%%%%%%%%%%%%%%%%%%%%%%%%%%%%%%%%%%%
%
% Modeling the EBL
%
%%%%%%%%%%%%%%%%%%%%%%%%%%%%%%%%%%%%%%%%%%%%%%%%%%%%%%

\section{Modeling the EBL}
\label{sec:eblmodeling}

We have used two methods for reconstructing the luminosity density and EBL from our $\g$-ray  optical-depth measurements.  The first is using the empirical method described by \citet{fermi2018gamma} and \citet{desai2019}.  Those authors applied this method to older $\g$-ray absorption optical depth measurements described in their papers.  The second is using the physically-motivated model of \citet{finke2022modeling}.  Previously those authors applied their model to the older $\g$-ray data.  Both of these model results are described below.
%\AD{[We should clarify that this latter procedure is also used in the previous papers Abdollahi+18 and Desai+19]}. \JB{[Maybe we also want to mention that the Finke model used in Abdollahi et al. (2018) and Desai et al. (2019) is the older version, while we're using the updated one in this work?]}

\subsection{Empirical Model}
\label{sec:empiricalmodel}
We first use the empirical model, fully described by \citet{fermi2018gamma} and \citet{desai2019}, and summarized below.  In this model, the luminosity density (also known as the emissivity) is fit with the sum of 7 log-normal distributions, 
\begin{flalign}
j_i(\lambda) = \sum_i a_i \exp\left[ -\frac{(\log_{10}\lambda - \log_{10}\lambda_i)^2}{2\sigma^2}\right]\ ,
\end{flalign}
at wavelengths $\lambda_i = [0.16, 0.50, 1.6, 5.0, 16, 50, 160]\ \mu$m.  Each component is allowed to evolve with redshift as
\begin{flalign}
\label{eqn:lumensevo}
j(\lambda_i, z) = j_0(\lambda_i) 
\begin{cases}
\frac{(1+z)^b_i}{1 + [(1+z)/c_i]^d_i} & i \le 3 \\
(1+z)^b_i & i >  3
\end{cases}
\ .
\end{flalign}
This model has free parameters $a_i$, $b_i$, $c_i$, and $d_i$ for each component.  The luminosity density is integrated over redshift $z$ to get the local ($z=0$) EBL intensity, 
\begin{flalign}
\label{eqn:EBLintensity}
\lambda I_\lambda = \frac{c}{4\pi}\int dz \frac{ \lambda j(\lambda/(1+z),z)}{(1+z)^2} \left|\frac{dt}{dz} \right|\ ,
\end{flalign}
where $|dt/dz|$ takes into account the expansion rate of the universe.  The EBL intensity as a function of redshift can be used together with the cross section for Breit-Wheeler pair production \citep[$\g + \g \rightarrow e^+ + e^-$; e.g.][]{gould67gg} to compute the absorption optical depth $\tau_{\g\g}(E)$ as a function of observed energy $E$.  This can be compared with the observed values of $\tau_{\g\g}(E)$ described above.

We performed an MCMC fit to {our} $\gamma$-ray {optical-depth} measurements, and the ones by \citet{baxter26} from VHE telescopes, with the empirical model described in this section using the {\tt emcee} routine \citep{foreman2013}.  This should be similar to the results of \citet{desai2019}, but using our updated $\g$-ray optical-depth measurements.  The $z=0$ EBL intensity results is shown in Figure \ref{fig:EBLintensity_empirical} with a variety of EBL measurements and models, including the result by \citet{desai2019}.  The results are comparable, although the intensity in our result are {lower}. The {\em New Horizons} measurement from \citet{postman2024new} is a bit high, but its 68\% uncertainty is still consistent with the empirical model. The Fermi-LAT energy range primarily probes the UV/optical portion of the EBL, and this is well-constrained.  In the IR region, between approximately 4 and 20 $\mu$m, our result is lower than the previous result by \citet{desai2019}.   This region is sensitive to the VHE optical depths, which have been newly derived in \citet{baxter26}.  In this $\g$-ray region the optical depth changes very little with energy, and thus it is difficult to constrain changes in the $\g$-ray optical depth.  This leads to greater uncertainty in the 4---20 $\mu$m EBL intensity. As we move to longer wavelengths still, the TeV data become more sparse, and the result becomes more uncertain.
%{\bf Justin: this needs to be rephrased: In the relevant $\g$-ray region the opacity changes very little with energy, and thus it is difficult to constrain changes in the $\g$-ray opacity.  This leads to greater uncertainty in the 4---20 $\mu$m EBL intensity. } 
Observations by CTA will be more constraining in this region, since it can observe out to higher energies with greater sensitivity.  At wavelengths longward of approximately 70 $\mu$m, the EBL intensity is not constrained by the $\g$-ray optical depth measurements at all, only by the EBL lower limits from \citet{driver2016measurements}.  The upper confidence interval is essentially unconstrained in this region.

\begin{figure}
%\vspace{2.2mm} \epsscale{1.0}
%\vspace{2.2mm} 
\epsscale{1.0}
\plotone{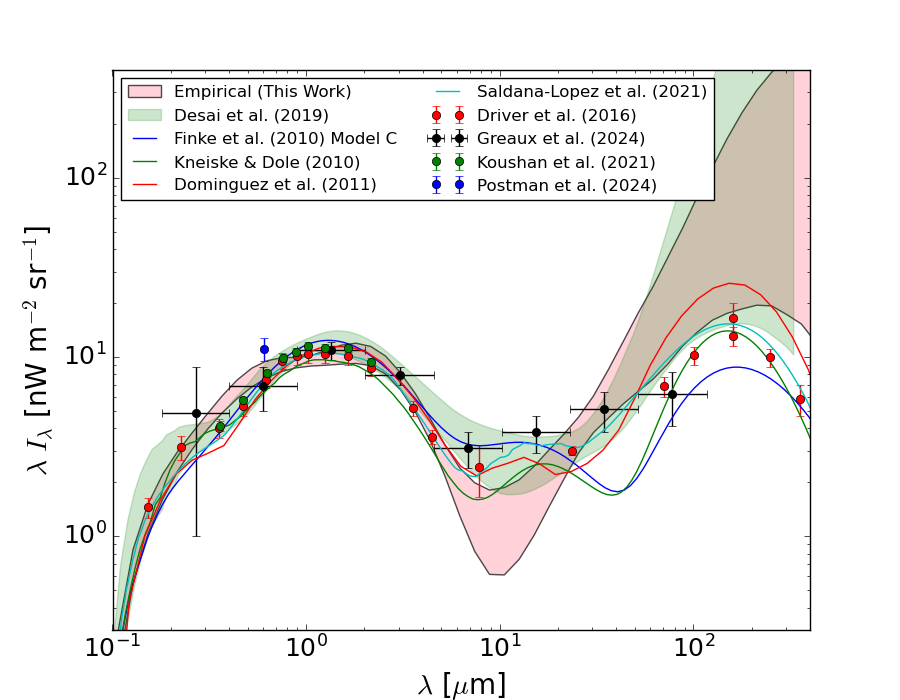}
\caption{68\% contour EBL intensity results as a function of wavelength at $z=0$ for our empirical model here.  Also shown are measurements and models, as described in the legend. }% \AD{[The Driver data should be updated with the more recent Koushan+21 that superseeded Driver, and update Biteau \& Williams for the updated Greaux+24 data. The legend should show only one symbol and not two symbols]} \JB{It would also be good to include one of the most recent direct measurement results in the plot, such as Postman et al. (2024).}}
\label{fig:EBLintensity_empirical}
%\vspace{-5.5mm}
\end{figure}

The luminosity density as a function of redshift $z$ at three wavelengths, UV (0.15 $\mu$m), optical (0.44 $\mu$m), and infrared (1.22 $\mu$m), for our empirical model result can be seen in Figure \ref{fig:lumdens_empirical}, along with a variety of measurements from galaxy surveys.  These data were compiled by \citet{finke2022modeling}. Our results are in good agreement with the observations, although they lie slightly below the measured values at 1.22 $\mu$m.  At redshifts $\ga 5$, the luminosity density is essentially unconstrained by our model, as the highest blazar in our sample is at $z=4.31$.  At redshifts higher than this, the only constraints come from our choice of functional form for the evolution of the luminosity density (Equation \ref{eqn:lumensevo}).

%{\bf\boldmath We have demonstrated that, using our $\g$-ray opacity measurements and some fairly basic assumptions, we can constrain the luminosity density of the universe out to $z\approx5$, when the Universe was a little more than a Gyr old.  Our measurement is in agreement with other, completely independent measurements from galaxy surveys.\unboldmath}
 We have demonstrated that, using our $\gamma$-ray opacity measurements and some fairly basic assumptions, we can constrain the luminosity density of the Universe out to $z \approx 5$, when it was a little more than a Gyr old. Our measurement is consistent with independent determinations from galaxy surveys, indicating that the empirical reconstruction captures the global evolution of the EBL and its underlying emissivity with sufficient accuracy to reproduce the main features observed in deep survey data.

\begin{figure}
%\vspace{2.2mm} \epsscale{1.0}
%\vspace{2.2mm} 
\epsscale{1.0}
\plotone{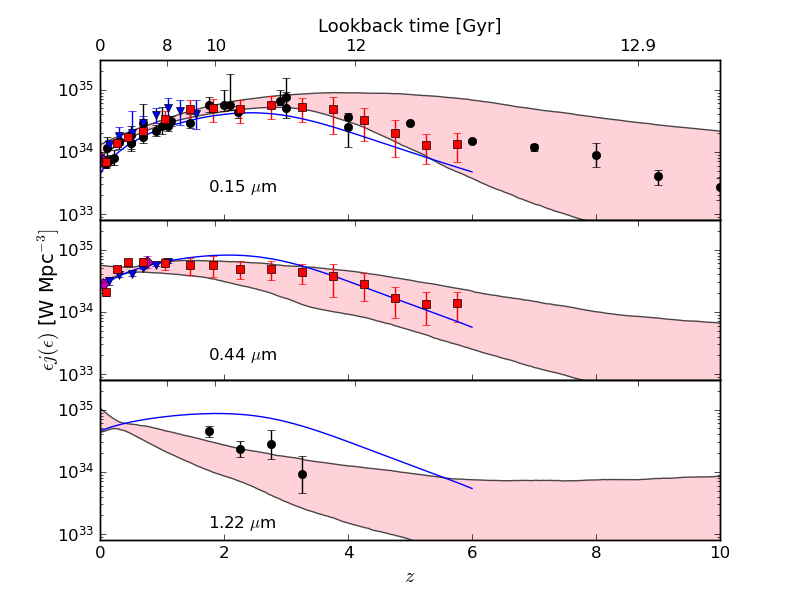}
\caption{68\% contour luminosity density results for as a function of redshift for three wavelengths (labeled on the plots) for our empirical model here.  
The blue curve is the model from \citet{finke2010modeling}, and the symbols show the measurements from \citet[][red squares]{saldana2021observational}, \citet[][blue triangles]{tresse2007}, and other sources (black circles) described in \citet{finke2022modeling}.}
\label{fig:lumdens_empirical}
%\vspace{-5.5mm}
\end{figure}

\subsection{Physically Motivated Model}

We next performed a similar MCMC fit to our $\gamma$-ray optical depth measurements, and the ones by \citet{baxter26} from VHE telescopes, with the physical model of \citet{finke2022modeling}. This physically-based model uses spectra from PEGASE, convolved with star formation, and takes into account dust extinction and the metallicity evolution of the universe.  As with the empirical model, we also fit the measurements from \citet{driver2016measurements} and \citet{koushan2021}, treating them as lower limits.  See \citet{finke2022modeling} for a full description of this model.  We took the four star formation rate parameters ($a_s$, $b_s$, $c_s$, and $d_s$) in an empirical function \citep[see][]{madau2014cosmic}, as well as $f_1$ and $f_2$, the fraction of absorbed light that is re-radiated in the infrared.  We assumed an initial mass function from \citet{baldry2003imf}.  Once the luminosity density is computed with this physical model, the model EBL intensity and $\g$-ray optical depth are computed as in the empirical model.
Since we are mainly fitting to the EBL $\g$-ray optical depth here, our physical model result here is most similar to Model B from \citet{finke2022modeling}, where earlier $\g$-ray optical depth data were fit.  The physically-based model is more constraining than the empirical model, since the expected results are also constrained by physics.  This model, as used here, only has 6 free parameters: four for the star formation rate ($a_s$, $b_s$, $c_s$, and $d_s$) and two representing the fraction of absorbed light re-radiated in two of the three dust components ($f_1$ and $f_2$).

The resulting model parameters are:  $a_s=0.025^{+0.008}_{-0.007}$, $b_s=1.5\pm0.5$, $c_s=4.0^{+1.1}_{-0.7}$, $d_s=7.0^{+1.9}_{-2.3}$, $f_1=0.4\pm0.3$, and $f_2=0.4\pm0.3$, which are completely consistent with the Model B result from \citet{finke2022modeling}.  The EBL intensity for our result here can be seen in Figure \ref{fig:EBLintensity_physical}.  Agreement is generally good with all other models and measurements shown here. As with the empirical model, the physically-motivated model is consistent with the {\em New Horizons} measurement \citep{postman2024new}. The far-IR component (peaking around $100\ \mu$m) is higher than the models from \citet{finke2022modeling}, \citet{finke2010modeling}, \citet{koushan2021}, and \citet{saldana2021observational}.   The uncertainties here are generally about the same as Model B from \citet{finke2022modeling}.

\begin{figure}
%\vspace{2.2mm} \epsscale{1.0}
%\vspace{2.2mm} 
\epsscale{1.0}
\plotone{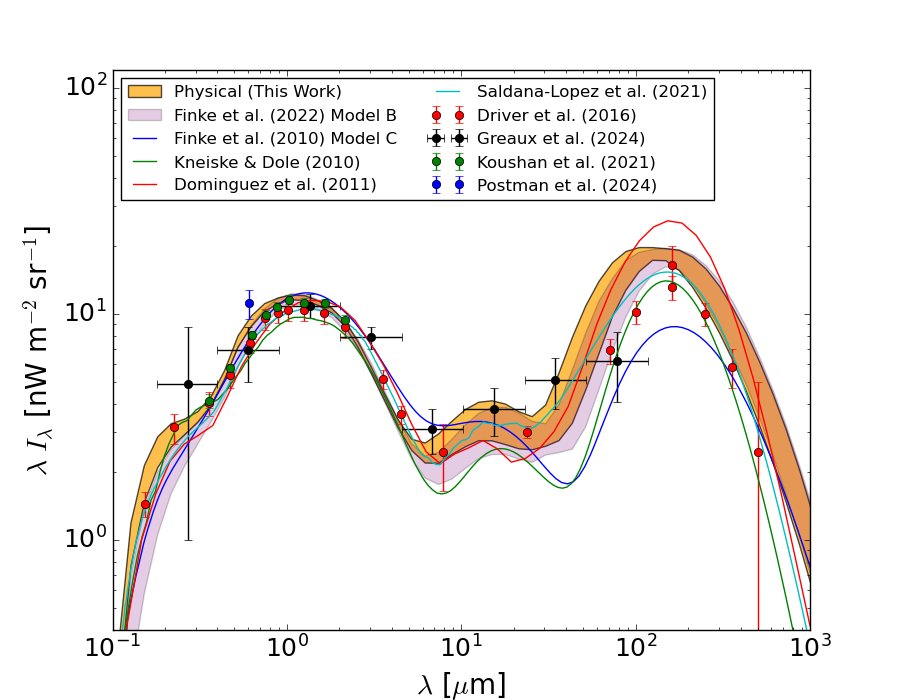}
\caption{68\% contour EBL intensity results as a function of wavelength at $z=0$ for our physically-based model here.  Also shown are measurements and models, as described in the legend.}
\label{fig:EBLintensity_physical}
%\vspace{-5.5mm}
\end{figure}

In Figure \ref{fig:lumdens} we show our model resulting luminosity densities as a function of redshift at three wavelengths:  UV (0.15 $\mu$m), optical (0.44 $\mu$m), and infrared (1.22 $\mu$m), along with other measurements from galaxy surveys, taken from the literature.  Our results are in agreement with these measurements and very similar to the previous result from \citet{finke2022modeling}.  They are also mostly in agreement with the older model from \citet{finke2010modeling}, except in the infrared.  

\begin{figure}
%\vspace{2.2mm} \epsscale{1.0}
%\vspace{2.2mm} 
\epsscale{1.0}
\plotone{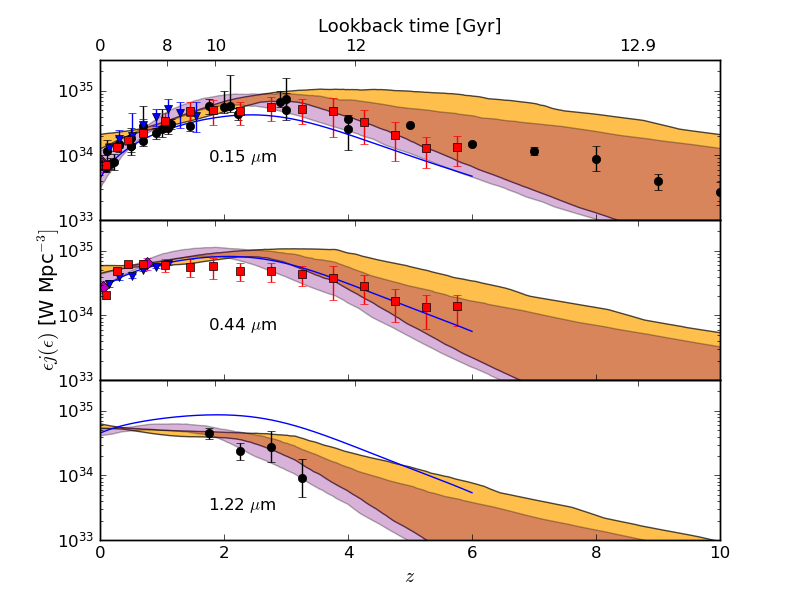}
\caption{The orange region shows the 68\% contour luminosity density results as a function of redshift for three wavelengths (labeled on the plots) for our physically-based model here.  The violet shaded region is the 68\% contour result from \citet{finke2022modeling} Model B.  The blue curve is the model from \citet{finke2010modeling}, and the symbols show the measurements from \citet[][red squares]{saldana2021observational}, \citet[][blue triangles]{tresse2007}, and other sources (black circles) described in \citet{finke2022modeling}.}
\label{fig:lumdens}
%\vspace{-5.5mm}
\end{figure}

In Figure \ref{fig:lumdens_staticz} we present our model resulting luminosity densities as a function of wavelength for different redshifts, along with various measurements. Our results are mostly in agreement with the luminosity density data.   At $z=0.10$, our results in the 0.1---1 $\mu$m region are high compared to the data, and the previous model; and at $z=0.90$, the model is below the data in the 0.4---3.0 $\mu$m region.  They are above Model B from \citet{finke2022modeling} due to the inclusion of the lower limit constraints from \citet{koushan2021}. 

\begin{figure}
%\vspace{2.2mm} \epsscale{1.0}
%\vspace{2.2mm} 
\epsscale{1.0}
\plotone{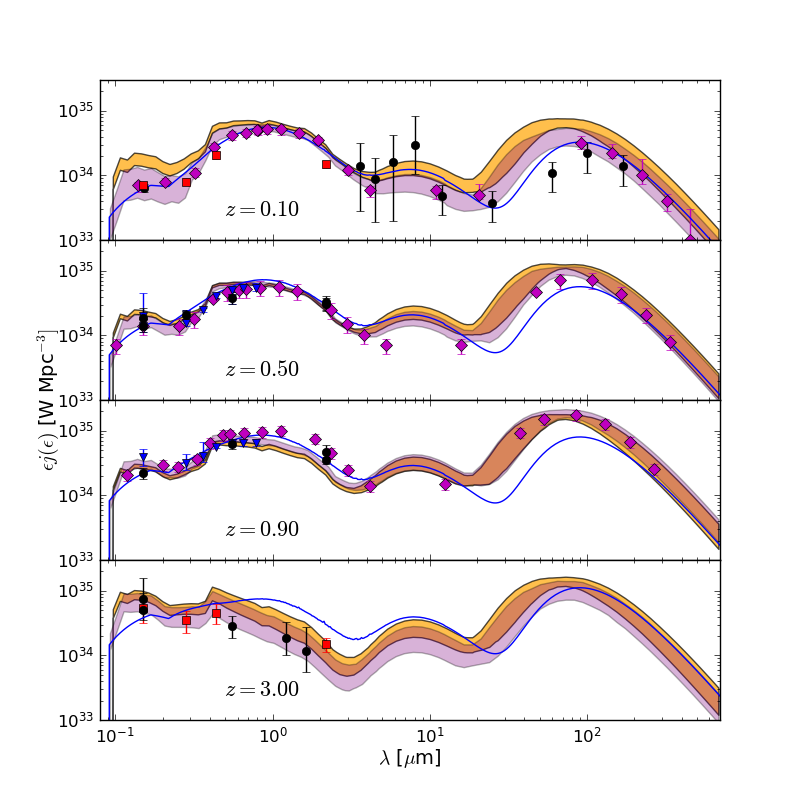}
\caption{The orange region shows the 68\% contour luminosity density results as a function of wavelength for four redshifts for our physically-based model here; the redshifts ($z$) are labeled on the plots.  The violet shaded region is the 68\% contour result from \citet{finke2022modeling} Model B.  The blue curve is the model from \citet{finke2010modeling}, and the symbols are the same measurements in Figure \ref{fig:lumdens}; additionally, the purple diamonds indicate the data from \citet{andrews2017}.}
\label{fig:lumdens_staticz}
%\vspace{-5.5mm}
\end{figure}

In Figure \ref{fig:SFRrate} we show the cosmic star formation rate as a function of redshift from the physical model presented here, and ``Model A'' and ``Model B'' from \citet{finke2022modeling}.  Their Model A included fits to a large amount of luminosity density data, in addition to $\gamma$-ray optical depth data; their Model B was a fit to the $\gamma$-ray optical depth data only.  Also shown are star formation rate data from 1500\ \AA\ luminosity density data. The conversion from luminosity density to star formation rate was done using the median values for the physical model in this work. The results are all in agreement, with out most recent star formation rate being higher at high $z$ than the previous Model B (which is the most comparable model from \citet{finke2022modeling}), but still consistent with it.  Uncertainties are quite large at $z\ga5$, since there are no $\gamma$-ray optical depth data at these high redshifts.

\begin{figure}
%\vspace{2.2mm} \epsscale{1.0}
%\vspace{2.2mm} 
\epsscale{1.0}
\plotone{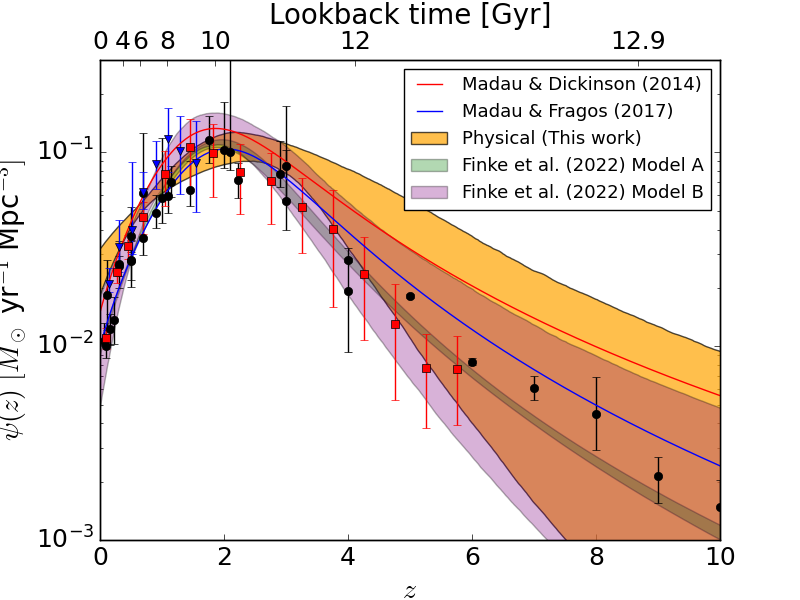}
\caption{68\% contour star formation rate density as a function of density for the physical model result presented here, and Model A (including a fit to luminosity density and $\gamma$-ray optical depth data) and Model B (which only did a fit to $\gamma$-ray optical depth data), and the curves from \citet{madau2014cosmic} and \citet{madau2017}.  The symbols are the same measurements in Figure \ref{fig:lumdens}.}
\label{fig:SFRrate}
%\vspace{-5.5mm}
\end{figure}

Finally, we ran a version of this model with the UV escape fraction, $f_{\rm esc,UV}$ as a free parameter.  This parameter controls the fraction of light that can ionize hydrogen (energy $>13.6$\ eV or wavelength $<912$\ \AA) and is important for reionizing the hydrogen in the universe \citep[e.g.,][]{munoz24}.  It has been suggested that $\g$-ray absorption measurements could constrain this parameter \citep{malkan21}.  We found this was not the case with our $\g$-ray data and EBL model \citep{finke2022modeling}.  Our result is that $f_{\rm esc,UV}$ is essentially unconstrained. \citet{malkan21} neglected the effect of dust extinction in their calculations, while \citet{finke2022modeling} included it by extrapolating from longer wavelengths down to below 912\ \AA.  This is likely the difference in our result and that of \citet{malkan21}.

%EBL photons with energy $\ge 13.6$\ eV should interact with $\g$-ray photons of energy $\le 19/(1+z)$\ GeV \citep[see, e.g., ][Equation (1)]{finke2022modeling}.  An inspection of Figure \ref{fig:tau} and Table \ref{tab:tau} indicates that this region is not measured by the data.  This is because the absorption is too small to be constrained at these energies.  Our result is that $f_{\rm esc,UV}$ is essentially unconstrained.

%{\bf\boldmath The combination of $\g$-ray measurements and physical modeling allows us to constrain the luminosity and star and formation history our to $z\approx 5$, when the Universe was a little more than 1 Gyr old.  Adding physical modeling allows for stronger constraints than the empirical modeling alone (Section \ref{sec:empiricalmodel}).  The results are in agreement with other, independent measurements from galaxy surveys.  \unboldmath}

 The combination of $\gamma$-ray measurements and physical modeling allows us to constrain the luminosity and star-formation history out to $z \approx 5$, when the Universe was a little more than 1~Gyr old. Adding physical modeling yields stronger constraints than the empirical reconstruction alone (Section~\ref{sec:empiricalmodel}), and the results remain in agreement with independent measurements from galaxy surveys. This consistency indicates that the $\gamma$-ray--driven constraints are compatible with standard assumptions about star formation and dust reprocessing.

%%%%%%%%%%%%%%%%%%%%%%%%%%%%%%%%%%%%%%%%%%%%%%%%%%%%%%%%%%%%%%%%%%%
%
%
%
% DISCUSSION AND CONCLUSIONS
%
%
%
%%%%%%%%%%%%%%%%%%%%%%%%%%%%%%%%%%%%%%%%%%%%%%%%%%%%%%%%%%%%%%%%%%%%%
\section{Discussion and Conclusions}
\label{sec:discussion}
The attenuation produced by the interaction of gamma-ray and EBL photons produces an imprint in the spectra of extragalactic sources. This attenuation has been successfully used in the past to measure the intensity of the EBL. Taking advantage of $\sim$15\,years of observations by {\it Fermi}-LAT and $>$1500 blazars, this work produces an updated measurement of the EBL optical depth over the $0<z<4.3$ range. This work employs the same analysis method of \cite{Abdollahi2018}, but relies on a 80\,\% larger {telescope time} and a blazar sample that has almost doubled. The attenuation of the EBL is collectively detected in the spectra of blazars {with $TS$ between 538 and 543, corresponding to $\sim23$\,$\sigma$. The $TS_{EBL}$ increased by 80\,\% between this analysis and \cite{Abdollahi2018}, in agreement with the increased exposure, indicating that the impact of newly detected, faint blazars is rather marginal. The increased significance of the EBL detection allowed us to measure the optical depth at 19 different epochs, up to $z\approx4.3$. At each epoch, most of the sensitivity of our measurement is found around the $\tau_{\gamma\gamma}=1$ transition from a transparent to an opaque Universe. This allowed us to measure the CGHR with unprecedented accuracy (see Figure~\ref{fig:cghr}).  These results stregthen the use of gamma-ray attenuation as a cosmological tool, in particular for constraining star formation at early epochs and for determining the Hubble constant through independent methods. In this context, the refined CGRH measurement provides a stringent observational benchmark for theoretical and empirical EBL models and represents the most accurate determination to date. It will also serve as a valuable reference for future observations by the Cherenkov Telescope Array Observatory (CTAO), which will explore gamma-ray energies up to the multi-TeV range, corresponding to the lower-redshift regime, with unprecedented sensitivity. This energy domain is currently probed by Fermi-LAT with larger uncertainties due to its limited sensitivity above a few hundred GeV.

We have used the GeV optical depths in Figure~\ref{fig:tau} and the TeV optical depths recently derived by \citet{baxter26} to deconvolve the intensity and the evolution of the EBL. This was done in two different ways: using an empirical reconstruction method \citep[described in][]{Abdollahi2018} and a physically-motivated one \citep[using the method of][]{finke2022modeling}.  {We are able to reconstruct the luminosity density and star formation rate density with reasonable accuracy out to $z\approx5$.  Our results are generally in agreement with previous measurements from gamma-ray optical depths and galaxy surveys.} Previously in \cite{Abdollahi2018}, the high-$z$ EBL constraints were entirely driven by GRB~080916C ($z=4.31$). This work only relies on blazars that reach the same redshift.

The good agreement between our reconstructed EBL and the IGL also enables us to investigate whether additional diffuse components may contribute to the cosmic optical background. A particularly relevant possibility is intra-halo light (IHL), produced by stars that have been tidally stripped and dispersed into the halos of galaxies at low redshift \citep{Zemcov2013, Zemcov2017}. If present, such emission would add a faint, spatially extended component to the EBL that may not be fully captured by current galaxy counts. To assess this scenario, we introduce a fractional scaling parameter $\alpha_{\mathrm{IGL}}$ that quantifies the strength of any diffuse component relative to the IGL, and we test for such a contribution by assuming that it follows the same spectral shape as the IGL. We derive a 95\% upper limit of $\alpha_{\mathrm{IGL}} < 0.23,$ indicating that any diffuse IHL-like component can contribute no more than approximately 23\% of the local EBL intensity. We note that this upper limit is consistent with recent TeV-based EBL reconstructions \citep{greaux24}, despite differences in the underlying data sets and analysis methods. This result is consistent with the view that the bulk of the EBL originates in resolved galaxy populations. It also mildly disfavors recent claims that diffuse IHL may contribute up to 30\% of the near-infrared background \citep{driver2016measurements, Cheng2022}.

The detection and characterization of the EBL using the gamma-ray absorption technique experienced a tremendous improvement in the last 15\,years thanks to observations in space by {\it Fermi}-LAT and on the ground by air-Cherenkov TeV telescopes. This technique has reached full maturity, and in the GeV energy range (corresponding to UV and optical EBL wavelengths), the results have reached an incremental stage. Assuming {\it Fermi} will remain operative for the next 10\,yr, we may expect another 80\,\% improvement in the $TS_{EBL}$, which may allow us to measure how the optical depth varies across 40 different epochs. The most important frontier, however, would be to extend this measurement to $z\ga5$, where constraining the UV luminosity density becomes important to understand the modes and times of re-ionization. To make substantial progress on this front, one would need a set of high-$z$ GRBs as bright as GRB 080916C ($z=4.35$), which remains to this day the only one of its kind.  Another dramatic improvement would be to provide redshift for all the BL Lacs detected by {\it Fermi}-LAT. Because of their hard GeV spectra, BL Lacs provide nearly 4 times the signal-to-noise ratio of FSRQs (see Figure~\ref{fig:stacked}). The 4LAC-DR3 high-latitude catalog contains 1379 BL Lacs and 1208 blazars of uncertain type, the majority of which are likely BL Lacs. Among these 2587 blazars, only 994 have redshifts. The rest, 1593 objects, do not yet have a redshift estimate. 
Photometric redshift programs \citep{rau2012,rajagopal2020,sheng2024,sheng2025} are able to discover hard-spectrum BL Lacs at $z>1$ at a rate of $\sim$10\,\% of the total BL Lac population. There could potentially be $\approx$150, $z>1$, BL Lacs in the 4LAC sample already. Measuring their redshift, and even better, measuring the redshift of all BL Lacs, would dramatically improve the characterization of the EBL using gamma rays.

\begin{acknowledgements}

{The Fermi-LAT Collaboration acknowledges generous ongoing support from a number of agencies and institutes that have supported both the development and the operation of the LAT as well as scientific data analysis. These include the National Aeronautics and Space Administration and the Department of Energy in the United States, the Commissariat \`a l'Energie Atomique and the Centre National de la Recherche
Scientifique/Institut National de Physique Nucl\'eaire et de
Physique des Particules in France, the Agenzia Spaziale
Italiana and the Istituto Nazionale di Fisica Nucleare in Italy,
the Ministry of Education, Culture, Sports, Science and
Technology (MEXT), High Energy Accelerator Research
Organization (KEK) and Japan Aerospace Exploration Agency
(JAXA) in Japan, and the K.\ A.\ Wallenberg Foundation, the
Swedish Research Council and the Swedish National Space
Board in Sweden.}

{Additional support for science analysis during the operations
phase is gratefully acknowledged from the Istituto Nazionale di
Astrofisica in Italy and the Centre National d’\`Etudes Spatiales
in France. This work performed in part under DOE contract
DE-AC02-76SF00515.}

A.B. and M.A. aknowledge NASA funding under contract 80NSSC22K1579.  J.D.F.\ was supported by NASA through contract S-15633Y; the Office of Naval Research; and by a grant of computer time from the Department of Defense High Performance
Computing Modernization Program at the Naval Research Laboratory. {A.\ Dom\'{i}nguez is thankful for the support of Proyecto PID2021-126536OA-I00 funded by MCIN / AEI / 10.13039/501100011033.} A.\ Desai was supported by an appointment
to the NASA Postdoctoral Program at NASA Goddard Space Flight Center, administered by Oak Ridge Associated Universities under contract with NASA.
J.B. gratefully acknowledges support from JSPS KAKENHI Grant No. 24KJ0545 and the JSPS Overseas Challenge Program for Young Researchers. J.B. also thanks UCM for hosting his six-month stay under the latter program, and A. Dom\'{i}nguez for his warm welcome.

\end{acknowledgements}

\bibliography{references}{}
\bibliographystyle{aasjournal}

%% This command is needed to show the entire author+affiliation list when
%% the collaboration and author truncation commands are used.  It has to
%% go at the end of the manuscript.
%\allauthors

%% Include this line if you are using the \added, \replaced, \deleted
%% commands to see a summary list of all changes at the end of the article.
%\listofchanges

\end{document}